\documentclass[fleqn,usenatbib]{mnras}

\usepackage{mathptmx}

\usepackage[T1]{fontenc}

\DeclareRobustCommand{\VAN}[3]{#2}
\let\VANthebibliography\thebibliography
\def\thebibliography{\DeclareRobustCommand{\VAN}[3]{##3}\VANthebibliography}

\usepackage{graphicx}	
\usepackage{amsmath}	
\usepackage{amssymb}	
\usepackage{url}
\usepackage{orcidlink}

\title[Comparing G inside and outside cosmic voids]{Using Elliptical Galaxy Kinematics to Compare of the Strength of Gravity in Cosmological Regions of Differing Gravitational Potential-- A First Look}

\author[E. Pedersen et al.]{
Eske M. Pedersen\orcidlink{0000-0002-8883-2172}$^{1}$\thanks{E-mail: eskepedersen@fas.harvard.edu},
Christopher W. Stubbs \orcidlink{0000-0003-0347-1724} $^{1,2}$
\\
$^{1}$Department of Physics, Harvard University, 17 Oxford street, Cambridge, MA 02143, USA\\
$^{2}$Department of Astronomy, Harvard University, and Center for Astrophysics, 60 Garden Street, Cambridge, MA 02143, USA\\
}

\date{Accepted XXX. Received YYY; in original form ZZZ}

\pubyear{2023}

\begin{document}
\label{firstpage}
\pagerange{\pageref{firstpage}--\pageref{lastpage}}
\maketitle

\begin{abstract}
Various models of modified gravity invoke ``screening'' mechanisms that are sensitive to the value of the local gravitational potential.  This could have observable consequences for galaxies. These consequences might be seen by comparing two proxies for galaxy mass -- their luminosity and their internal kinematics -- as a function of local galaxy density. Motivated by this prospect, we have compared the observed properties of luminous red galaxies (LRGs) inside and outside of voids in the cosmic large scale structure. We used archival measurements of line widths, luminosities, redshifts, colors, and positions of galaxies in conjunction with recent void catalogs to construct comparison LRG samples inside and outside of voids. We fitted these two samples to the well-established fundamental plane of elliptical galaxies to constrain any differences between the inferred value of the Newtonian gravitational constant G for the two samples. We obtained a null result, with an upper limit on any fractional difference in G within and outside of cosmological voids to be $\alpha =\delta$$ G/G \sim$ 40\%.  This upper bound is dominated by the small-number statistics of our N $\sim $ 100 within-void LRG sample. With the caveat that environmental effects could influence various parameters such as star formation, we estimate that a 1\% statistical limit on $\alpha$ could be attained with data from 10${^5}$ elliptical galaxies within voids. This is within the reach of future photometric and spectroscopic surveys, both of which are required to pursue this method. 
\end{abstract}


\section{Introduction}

There is a long history \citep{Skordis:2019fxt} of attempting to circumvent the need for dark matter by considering modifications to general relativity (GR). There has yet to be a single modified gravity model that accounts for all the observational data without appealing to dark matter. These observational data include rotation curves and velocity dispersions of galaxies, the growth of the large scale structure, gravitational lensing systems, and the discrepancy between $\Omega_{Baryon}$ and $\Omega_{mass}$ seen in the angular power spectrum of the cosmic microwave background (CMB) fluctuations, and other cosmological observations.  

The discovery of the accelerating expansion of the Universe \citep{riess1998observational,perlmutter1999measurements} and its interpretation as dark energy have provided added impetus to asking how well we understand gravitation.
Reviews from this perspective are provided by \cite{koyama2016cosmological} and \cite{baker2015linking} who emphasize the importance of testing gravitational physics in a 2-D space comprising local curvature and gravitational potential. 

In order to retain consistency with existing limits on extensions to GR, especially precision tests in the solar system, some of these proposed extensions invoke a `screening'' mechanism that suppresses modified gravity phenomenology in various ways. One complication that arises in scenarios that invoke screening is proper consideration of the various contributions to the local potential; cosmological background, the local galaxy, the solar system, and laboratory apparatus. 

Our current lack of understanding of both dark matter and dark energy motivates testing gravitation on all accessible length and energy scales. In this paper we establish a method and present initial observational limits on the extent of ``environmental screening'', due to the background gravitational potential, experienced by galaxies inside and outside of cosmic voids. 

We use the following cosmological parameters throughout: $h$ = 0.719, $\Omega_{\Lambda}$ = 0.742, $\Omega_m$ = 0.258. 

\subsection{Screening theories and observables}

Many modified gravity theories rely on some sort of screening to evade the constraints put on it by the solar system and other tests of general relativity \citep{Joyce_2015, DESC_MG, Ishak_2018, brax2021testing}. These screening models usually couple to the underlying density field/Newtonian potential/gravitational potential in one way or another. There are several proposed models of how this screening might couple to the gravitational potential: directly (Chameleon) or to its derivative (Vainshtein, K-mouflage) \citep{brax2021testing}. These screening models have already undergone several tests and limits have been put on them by both laboratory and solar system tests. Existing and proposed astrophysical tests \citep{Joyce_2015, DESC_MG, Ishak_2018, brax2021testing, cabre2012astrophysical, hui2009equivalence} can also provide interesting limits on longer length scales.  A recent paper by \cite{cataldi2022fingerprints} compared the shapes of spiral galaxy halos inside and outside of voids in numerical simulations, and is complementary to our approach. 


The hierarchy of mass concentrations in the cosmos determines the depth of the gravitational potential that embedded objects experience. Stars are within galaxies, which are themselves embedded in peaks and troughs in the large-scale structure density field. In the project described here, we assess the extent to which the local cosmic density, averaged on $\sim$ 10 Megaparsec scales (the typical extent of the voids used here), might influence the value of the Newtonian gravitational constant G that would cause a difference between astronomical observables of galaxies that lie inside and outside of voids. 

An interesting approach is to use observations of galaxy density to generate ``screening maps'' \citep{shao2019screening, cabre2012astrophysical} of a scalar quantity that characterizes the local strength of any proposed modification to GR. 
Cosmic voids are extensive sparse volumes that have galaxy number counts a factor ten below the cosmic mean. From the context of exploring the potential impact that the environmental background potential might have on modifications to G, it is natural to compare gravitational phenomena inside and outside of voids. 





There are multiple ways for modified gravity to impact observables of galaxies within voids. Here we explore how the internal kinematics of elliptical galaxies depend on the value of G that determines their binding energy. At fixed galaxy mass larger values of G will increase a galaxy's velocity dispersion. 

Another possible observable consequence of a non-standard value of G is on stellar evolution. The energy balance between the fusion and gravitational processes that govern luminosity evolution will depart from canonical values, leading to changes in the evolution of the apparent mass-to-luminosity ratio of a galaxy. Stars burn faster and hotter if G is increased \citep{davis2012modified}. This would modify the galaxy's luminosity as compared to its velocity dispersion. 

Finally, void evolution and peculiar velocities could also be influenced by screening mechanisms that modify G \citep{cautun2018santiago}, which is beyond the scope of this paper. 


Our approach is to use luminosity as a proxy for galaxy mass. Line widths in spectroscopic absorption features are used to determine the velocity dispersion for the galaxies. We look for inconsistencies between the mass estimated by starlight and the mass implied by the velocity dispersion, by comparing samples inside and outside of cosmic voids. This allows us to place limits on any discrepancy in the values of G inside and outside of voids. 

Choosing Luminous Red Galaxies (LRGs) as tracers for any variation in G has a number of advantages. The signal-to-noise ratio for photometric and spectroscopic measurements of the properties of these objects is high due to their intrinsic brightness. Compared to spiral galaxies, inclination effects are less impactful. There is less dust, and consequently less photometric attenuation, due to extinction within LRGs. 

\subsection{Fundamental plane of elliptical galaxies}

The subset of elliptical galaxies referred to as Luminous Red Galaxies (LRGs) have been shown to have a relation between observables of the galaxy \citep{Jorgensen}, specifically the velocity dispersion $(\sigma)$, the effective radius $(R_e)$ and the average surface brightness $(\langle I \rangle)$. Traditionally this relationship is expressed in their logarithmic versions called the fundamental plane 
\citep{Jorgensen, mo_van_den_bosch_white_2010}:
\begin{equation}
    \label{eq:fp}
    \log_{10} R_e = a \log_{10}\sigma + b \log_{10} \langle I\rangle + c,
\end{equation}
with $a$,$b$, and $c$ being fit parameters. As suggested in \cite{mo_van_den_bosch_white_2010} the fundamental plane can be related to the Virial theorem:
\begin{equation}
    \label{eq:vt}
    \frac{GM}{\langle R\rangle} = \langle v^2\rangle,
\end{equation}
where $M$ is the mass of the galaxy, $\langle R \rangle$ is the average radius, and $\langle v^2 \rangle$ is the average squared velocity. These can all be related to the observables of the fundamental plane:
\begin{align}
    R_e &= \mathcal{C} \sigma^2 \langle I \rangle^{-1} \left( \frac{M}{L}\right)^{-1}  \label{eq:Re}\\
    R_e &= k_R \langle R\rangle \\
    \sigma &= k_v \sqrt{\langle v^2 \rangle}\\
    L &= 2\pi \langle I\rangle R_e^2 \\
    \mathcal{C} &= \frac{1}{2\pi G k_R k_v^2}, 
\end{align}
with $k_v$, $k_R$ being scale factors.

The terms in the Virial theorem (Equation \ref{eq:vt}) are related to the terms in the fundamental plane, with Equation \ref{eq:fp} obtained by taking the logarithm and making appropriate substitutions. From this we see that the constant $c$ in Equation \ref{eq:fp} will be proportional to $\mathcal{C}$. More importantly for our case, $c$ will be proportional to $-\log_{10} G$.

From the Virial theorem shown in Equation \ref{eq:vt} we would expect $a = 2$ and $b=-1$ for the fit to the fundamental plane in Equation (\ref{eq:fp}), as long as all galaxies have the same mass-to-light ratio. However \cite{Jorgensen} found empirical values of $a \approx 1.2$ and $b\approx-0.8$. See the discussions in \cite{Jorgensen, mo_van_den_bosch_white_2010} for more details.


For observations that span a non-trivial redshift range, which is the case for the LRGs studied here, we also need to account for any redshift dependence in the observed quantities. The redshift of an LRG impacts its apparent surface brightness in three main ways: 
\begin{enumerate}
    \item Cosmological surface brightness dimming induces a reduction in surface brightness proportional to $(1+z)^4$ 
    \item The evolution of galaxy luminosity as we look back to earlier epochs, and
    \item In a given terrestrial photometric passband, we are integrating over increasingly bluer regions of the galaxy spectrum as redshift increases. In the lexicon of the photometry of extragalactic sources this is often termed a K-correction. 
\end{enumerate}



We therefore allow for a general power-law redshift dependence in our fitting methodology to account for effects beyond just modification to the apparent surface brightness. This adds a redshift-dependent additive term to the logarithmic fundamental plane relation. 

\subsection{A differential comparison of values of G}
\label{sec:Galpha}
We will make a simple phenomenological parameterization where the effective value of G inside a cosmic void takes on a modified value of $G_\text{mod} = G(1+\alpha)$. Here $\alpha$ is a dimensionless number that is expected to be zero outside of voids. The fundamental plane fitting parameter $c$ was shown to be proportional to $-\log_{10} (G_\text{effective})$, which implies that we can associate any difference between values of $c$ in and out of voids with:

\begin{equation}
\label{eq:deltac}
\Delta c = c_\text{non-void} - c_\text{void} = log10(1+\alpha),    
\end{equation}
and so 
\begin{equation}
\label{eq:alpha}
    \alpha = 10^{\Delta c}-1.
\end{equation}

\noindent
where $\Delta c$ is the difference in fundamental plane fit parameters for the in-void vs. out-of-void galaxy samples. 

\subsection{Structure of this paper}

We first describe the selection of both voids and of LRGs, and how we assign these galaxies to being inside or outside voids. The observable galaxy attributes of luminosity (using SDSS spectra to convert redshifts to distances and thereby from observed magnitudes to rest-frame luminosities) and velocity dispersion (taken as a proxy for mass in a virialized system) are then compared for the in-void and out-of-void populations.  

\section{Catalogs}
We use two catalogs: a catalog of void centers based on the SDSS data-release 7 \citep{Douglass:2022ebt}; and the summary catalog of the spectroscopic and photometric galaxy properties from SDSS dr17 \citep{SDSSDR17}. 

\subsection{Void catalog}
The void catalog we use is the VoidFinder based catalog described in \cite{Douglass:2022ebt}. This catalog found 1159 voids in the SDSS data release 7 using a redshift limit of $z\leq 0.114$ and required an absolute r-band magnitude of catalogued galaxies of $M_r \leq -20.$ The VoidFinder algorithm designates all the voids as spheres and finds that the voids in the sample have a median effective radius of $R_\text{eff} = 15.6\pm0.1 h^{-1}\text{Mpc}$. These voids contain $60.5\%$ of the survey volume but only $17.9\%$ of the galaxies. 

\subsection{LRG sample from SDSS-BOSS DR17}
\label{sec:spAll}
From the SDSS summary catalog (SpAll) \citep{SDSSDR17} catalog, we utilize the BOSS target selection to find the LRGs that are located within the redshift limit imposed by the void catalog. The photometric reduction pipeline provides us with the Petrosian angles and the average surface brightness as a flux within a series of apertures in each band. It also gives us an estimate of the galactic extinction for each object in each band. From the "spZbest" sub-catalog we get the redshift of each galaxy along with their associated velocity dispersion. 

\subsection{LRG properties}
\label{sec:LRGprop}
For the selected LRG galaxies we are interested in their redshifts, their photometric radii, their velocity dispersions, and their average surface brightness. 

\begin{figure}
    \centering
    \includegraphics[width=\columnwidth]{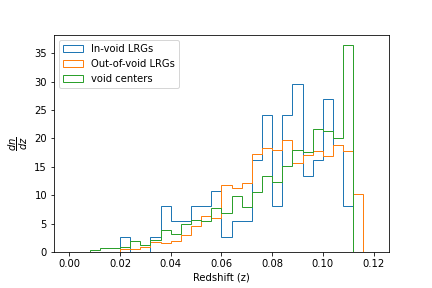}
    \caption{The redshift distribution of the LRG samples both the one in and out of voids. Also plotted is the redshift distribution of void centers, Each distribution is normalized such that the integral of the area under the graph is equal to 1. The samples contains 93 LRGs in the in-void sample, 3667 LRGs in the out-of-void sample, and 1159 voids.}
    \label{fig:Redshiftdist}
\end{figure}

Figure \ref{fig:Redshiftdist} shows the redshift distributions for the LRGs assigned to voids and the out-of-void sample. We see no substantial difference. We conducted a Kolmogorov-Smirnov test and found that we could not reject that the two samples are pulled from the same underlying distribution. This implies that spurious effects due to any unaccounted redshift dependence in LRG properties would be a second order effect in our differential comparison. We also overlapped the graph with the redshift distribution of the void centers. 

\begin{figure}
    \centering
    \includegraphics[width=\columnwidth]{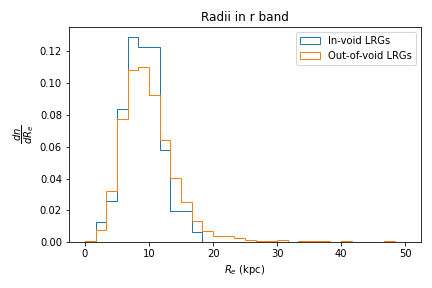}
    \caption{Shown here are the LRG radii determined from the $r$ band, the radii distributions for the other bands are shown in Appendix \ref{App:colordist}. Each distribution is normalized such that the integral of the area under the graph is equal to 1.}
    \label{fig:radii}
\end{figure}

Figure \ref{fig:radii} shows the effective radii distribution for both samples of LRG galaxies. To obtain the radii we start with the Petrosian angle $\theta_\text{petro}$ along with an angular diameter distance $d_A$ which can be obtained from our fiducial cosmology and the redshift of the galaxy:
\begin{equation}
    \ell = d_A \times \theta_\text{petro}.
\end{equation}
As with the redshift distribution, we also conducted a Kolmogorov-Smirnov test to determine the similarity between our two samples and found that the hypothesis that the two samples are drawn from the same distribution cannot be rejected. We illustrate here only the radii determined from the r band angle, but in Appendix \ref{App:colordist} we show the radii obtained in all bands of SDSS. 

\begin{figure}
    \centering
    \includegraphics[width=\columnwidth]{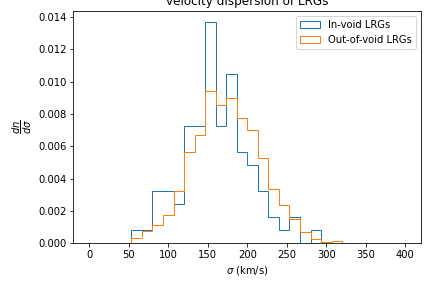}
    \caption{The velocity dispersion for the galaxies in both samples of LRGs. Each distribution is normalized such that the integral of the area under the graph is equal to 1.}
    \label{fig:vdisp}
\end{figure}

Figure \ref{fig:vdisp} compares the velocity dispersion for the two galaxy samples. The velocity dispersions plotted here are taken directly from the catalog described in Section \ref{sec:spAll}. We notice a slight shift between the two distributions, and the Kolmogorov-Smirnov test also suggests a slight discrepancy between the two samples, with LRGs outside of voids having on average a higher mass. This is presumably due to a higher merger rate outside of voids. 

Finally we compared the surface brightness for each galaxy. This is estimated in a set of circular apertures with fixed radii in arcsec for each object. We take the values of these in a spline and interpolate the value of the average surface brightness to the angle that corresponds to include $50 \%$ of the Petrosian flux of the object, this is the method suggested by \cite{SDSS-web}. The histogram for the r-band is shown in Figure \ref{fig:colorr}, while values for all bands is shown in Appendix \ref{App:colordist}. The flux here is given in nano-maggy per arcsecond squared, which is linear in surface brightness. 

\begin{figure}
    \centering
    \includegraphics[width=\columnwidth]{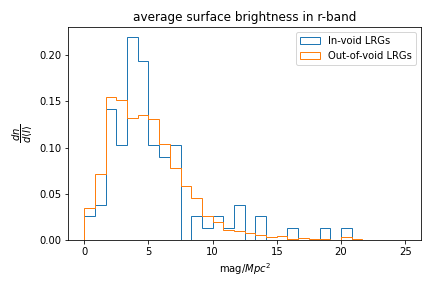}
    \caption{The r-band surface brightness histogram for the LRG samples.  We show surface brightness for the rest of the bands in Appendix \ref{App:colordist}.}
    \label{fig:colorr}
\end{figure}


\section{Method}

\subsection{Assigning galaxies as being inside voids}
\begin{figure}
    \includegraphics[width=\columnwidth]{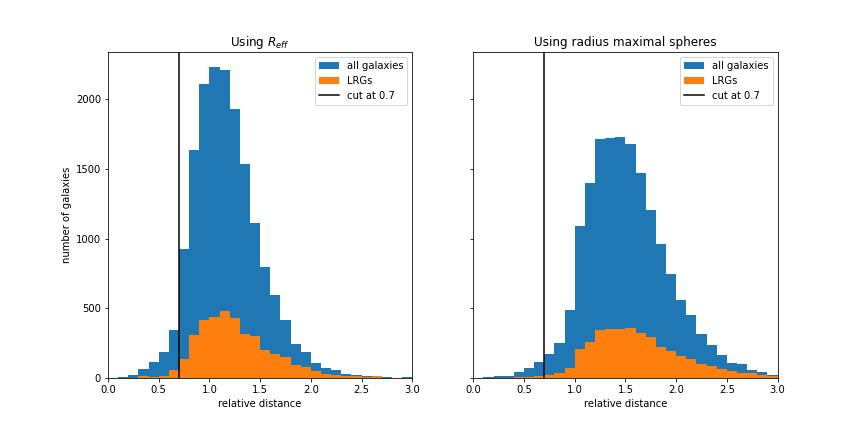}
    \caption{The distribution of relative distances between the galaxies and their closest void center, shown for both radii of the voids: left is the effective radii, right is the maximal sphere radii. Also shown is the distribution for luminous red galaxies (LRGs). We also show the cut-of value of 0.7 to distinguish which galaxies are in the voids and which ones are outside the voids. For the rest of this paper we will utilize the the effective radii measurements.}
    \label{fig:relativedist}
\end{figure}

We compared the 3-D locations of the galaxies to the positions and sizes of the catalogued voids; first we found which of the void centers are the closest to each galaxy, then we compared, for each galaxy, the distance to the closest void to the effective radius of that specific void. When we calculated the relative distance we also take into account any uncertainty in redshift. This left us with an easy way to estimate which galaxies are in-voids and which are out-of-voids.  
We showed the distribution of galaxies relative distance to their nearest void in Figure \ref{fig:relativedist}. Each void listed in \cite{Douglass:2022ebt} is assigned two different radii. The first is the effective radius of the combined/aggregated void and the second is the radius of the largest sphere which forms the center of the void. We choose to use the effective radii $(R_\text{eff})$ as depicted in the left hand side of Figure \ref{fig:relativedist}. To determine which galaxies constitute our in-void sample we choose a cut-of value at 0.7 in the relative distance to define our in-void sample. This leaves us with 735 galaxies within the catalogued voids;  93 of these are identified as LRGs. As described below, this sparse sample size limits the statistical significance of the upper limits we can place on any variation in G. 

\subsection{Data wrangling}
The different LRG properties we look for are described above, but to get these observables we also need to undertake a number of corrections:
\begin{itemize}
\item Correcting the photometry for galactic extinction, using the provided extinction values for each band, based on the work of \cite{1998SchlegelDavidFinkbeiner}. 
\item Converting redshifts to co-moving distance, this is done by calculating the angular diameter distance for each galaxy. We used $h$=0.70 to convert from redshifts to distances.  
\item Estimated the $\log_{10}$ values for the following versions of the observables described in Section \ref{sec:LRGprop}: $R_e$, $\sigma$, and $\langle I\rangle$\footnote{corrected for galactic extinction}. If any of these values were found to be not well defined i.e. resulting in infinities or NaN values, we omitted the entire galaxy from our sample. 
\end{itemize}
Besides these corrections, we imposed several cuts to the SDSS catalog:
\begin{itemize}
    \item cut in redshifts: $z>0.01$ and $z<0.114$
    \item Ensured there are no warnings on the redshift $(\text{'ZWARNING'} == 0)$ and $(\text{'ZWARNING\_NOQSO'} == 0)$.
    \item Required SDSS-BOSS identifier ('CLASS'=='GALAXY')
    \item Used the best observation of the spectra with ('SPECPRIMARY' == 1)
    \item cut in 112 degrees < RA < 259 degrees to coincide with the void catalog. 
\end{itemize}

With all the cuts listed here we end up with 17007(3785) galaxies(LRGs) in the general dataset. Of these, 735(93) galaxies(LRGs) are designated as in-void galaxies with the remaining 3667 as out-of-void LRGs. 

\subsection{Fundamental plane fitting}

For each LRG sub-sample of interest, we performed a double fit to Equation (\ref{eq:fp}) using the corrected observables for each LRG within the subset. First we did a least-square fit over the entire sample using SciPy's optimization package \citep{2020SciPy-NMeth}. We applied a robust loss function to exclude outliers. Having obtained an initial set of values for the fitting parameters we found for each galaxy the residual for $\log_{10} R_e$. These residuals we then compared for with the redshift of the galaxies. To this relation (residual/redshift) we fit a second order polynomial. The second order polynomial was used to tweak the galaxy's $\log_{10} R_e$ with a small correction, and the new value was used for a second curve fit to the fundamental plane. This gave a more robust set of fitting parameters $a,b,c$ and their variances. 

\begin{figure}
    \centering
    \includegraphics[width=\columnwidth]{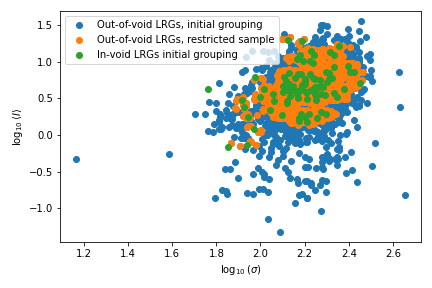}
    \caption{The fundamental plane projected to the $\log_{10} \sigma$/$\log_{10} \langle I\rangle$ plane of the restricted LRG sample. We picked out-of-void LRGs that are closer than the mean distance between out-of-void LRGs and their closest in-void LRGs.}
    \label{fig:restriction}
\end{figure}

To ensure that the fundamental plane fits are taken across similar regions in the observable space, we selected a restricted subset of the out-of-void LRGs.  We found the out-of-void LRG galaxies that lie close to the in-void LRGs in the plane spanned by $\log_{10} \sigma$ and $\log_{10} \langle I \rangle$. This is done by measuring the distances in this plane between the in-void LRGs and the out-of-void LRGs. We then did a cut that picked the out-of-void LRGs where this distance is below the mean distance in this space. This new sample of out-of-void LRGs, that we refer to as the restricted out-of-void LRG sample gives us a sample of LRGs outside voids that has observables similar to those of  in-void LRGs. We don't restrict in the $\log_{10} R_e$ and $z$ planes because these are corrected by the polynomial fit above. The restricted sample consists of 2658 LRGs. We then recalculated the fundamental plane fit on this restricted sample. The distances were estimated using SciPy's vq method. The resulting restriction in the $\log_{10} \sigma$ and $\log_{10} I$ plane is shown in Figure \ref{fig:restriction}. It is clear that the restricted sample is much closer to the in-void sample compared to the general out-of-void sample.

\subsection{Assessing statistical constraints on G}


To assess the statistical significance of any differences in the fundamental plane fits for the in-void and out-of-void LRG samples, we performed a bootstrap test using random draws of LRGs from the restricted non-void sample. We pulled 93 LRGs from the restricted sample at random. We then fitted the fundamental plane to each. We repeated this process 1000 times. Using these fit values of $c$, we estimated $\alpha$ using Equations \ref{eq:deltac} and \ref{eq:alpha} where we used the random draw in place of $c_\text{void}$. 

We also performed a validation test by injecting a multiplicative change in all velocity dispersions of the in-void galaxy sample. We verified that the resulting values of $\alpha$ were consistent with the corresponding artificial changes in G. 

\section{Results}

Table \ref{tab:results} summarizes the results obtained from the different fits. All samples here use r-band when relevant. 
\begin{table*}
    \centering
    \caption{Results for the fundamental plane fit to each of the samples. Coefficients for the fundamental plane parameters for the in-void, out-of-void both the restricted and non-restricted samples, as well as mean from the bootstrap tests are shown. The 1 $\sigma$ uncertainties are also shown. All values are estimated using the r-band values where relevant.}
    \label{tab:results}
    \begin{tabular}{|l|l|l|l}
\hline
Subset & a & b & c \\ 
\hline
In-void LRG sample & 0.930 $\pm$ 0.067 & -0.578 $\pm$ 0.030 & -0.732 $\pm$ 0.140 \\ 
Out-of-Void LRG sample & 0.953 $\pm$ 0.020 & -0.607 $\pm$ 0.008 & -0.781 $\pm$ 0.043 \\ 
Out-of-Void Restricted LRG sample & 0.940 $\pm$ 0.022 & -0.702 $\pm$ 0.010 & -0.678 $\pm$ 0.047 \\ 
Mean values for random replica & 0.931 $\pm$ 0.122 & -0.702 $\pm$ 0.053 & -0.657 $\pm$ 0.261 \\ 
\hline
    \end{tabular}
\end{table*}
 We see that the $a$ parameter is more or less unchanged throughout the samples. The $b$ and $c$ parameters do change, but primarily between the restricted sample and the non-restricted sample. The $b$ value jumps from around -0.6 to -0.7 while the $c$ goes from -0.78 to -0.68 when restricting the out-of-void sample. 

Of primary interest to us is constant $c$, which carries information about the effective value of $G$. 

\begin{figure}
    \centering
    \includegraphics[width=\columnwidth]{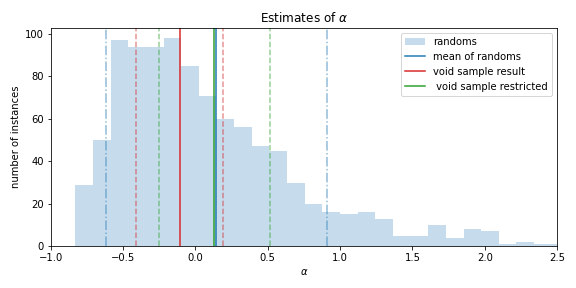}
    \caption{Histogram of the distribution of $\alpha$s for replacing the in-void-sample with the random pull comparing to the reduced out-of-void sample. Also shown is the mean value of these random pulls (blue) along with it's associated error bars (dashed-blue) (All error bars shown are the 1 $\sigma$ errors). We also show the values of $\alpha$ for the in-void sample vs the non-restricted (red) and the restricted out-of-void sample (green), both shown with error bars given by the least square fit (dashed red and dashed green respectively). }
    \label{fig:estimc}
\end{figure}




\subsection{Constraints on G}

The results of Table \ref{tab:results} show that the in-void and out-of-void LRG samples have a difference of $\Delta c= -0.0493(0.0536) \pm 0.146(0.148)$, at 95\% confidence for the unrestricted(restricted) samples. This implies a constraint of $\alpha = -0.107(0.131) \pm 0.301(0.384)$. 

\subsection{Potential astrophysical sources of systematic errors}

A variety of astrophysical factors could produce effects that mask or mimic the gravitational physics we are exploring here. One potential source of concern is that the observed properties of these galaxies could depend on the local galaxy density. for example, we have implicitly assumed that the mass-to-light ratio of these LRGs does not depend on the galaxy density in their local neighborhood. This assumption could be explored using N-body simulations, but that is beyond the scope of this paper.  

Figures \ref{fig:Redshiftdist}, \ref{fig:radii}, \ref{fig:vdisp}, \ref{fig:colorr}, \ref{fig:colors}, and \ref{fig:radiis} show comparisons of the properties of the LRGs.

Table \ref{tab:systematics} lists some of the astrophysical factors that merit further exploration.


\begin{table*}
    \centering
    \caption{Potential sources of astrophysical systematic error, and suggestions for exploring them in the future.}
    \label{tab:systematics}
    \begin{tabular}{c|c}
    \hline
    Source of Uncertainty &  Possible Approaches \\
    \hline
    Not volume complete LRG sample  & better detailed survey\\
     Environmental dependence of Dark Matter fraction in LRGs   & N-body simulations? \\
     Stellar evolution differences due to environment  &  metallicity and SFH modeling \\
     Misidentifying of void-galaxy & Forward modelling of density field to identify voids. \\
     Differences in Non-void LRGs dependent on their surroundings & Forward modelling of density field to distinguish clusters from sheets.\\
     
       \hline
    \end{tabular}
\end{table*}

\section{Conclusions}

Using the fundamental plane relationship for LRGs, we have performed a differential test for any variation in the effective value of G that acts on galaxies inside and outside of cosmic voids. The inside and outside of cosmic voids have different large-scale averaged values of the local gravitational potential, and this result sets a limit on modified gravity models in which gravitational physics could depend on these large-scale embedded gravitational potentials. 

We found no difference in the gravitational strength G for LRG kinematics inside and outside of voids. If G=G(1+$\alpha$) inside voids, we established a 1 $\sigma$ bound of $\|\alpha\|<0.4$.   

\subsection{Next steps and future work}

One refinement of this project would be to determine,  the local density on different spatial averaging scales for each LRG. This would be a more nuanced approach than simply declaring galaxies to be inside or outside of a void. An even better addition would be to estimate the local cosmological gravitational potential and a corresponding screening map for each proposed modified gravity theory, along the lines of \cite{Tsaprazi_2021, Jasche_2019}, and to seek any dependence of G$_\text{effective}$ as a function of various screening parameters. 

Another path of improvement would be to utilize a larger galaxy catalog that includes fainter elliptical galaxies drawn from a larger survey volume. The statistical power of this test is limited by the number of elliptical galaxies $N_{void}$ assigned to voids. Based on jackknife tests, we confirmed a good estimator for $\sigma_{\alpha}$ is 0.4 $\sqrt{(100/N_{void})}$. Achieving a statistical 1 $\sigma$ uncertainty in  $\alpha$ of 1\% implies $N_{void} \sim $ 1.6 $\times 10^5$. We stress that both photometric and spectroscopic information is needed for these galaxies. The DESI survey \citep{myers2023target} expects to measure the properties of 625 LRGs per square degree over one third of the sky. We should therefore expect DESI spectroscopic data for 8.5 million LRGs. If the DESI LRG fraction that resides in voids is comparable to what we see in the SDSS data, we should expect $2.5\%$ of DESI LRGs would satisfy the conservative in-void cut-off we imposed, yielding an expected $N_{void}$ from DESI on the order of 2 $\times$ 10$^5$. This implies that a 1\% statistical uncertainty in $\alpha$ should be attainable, subject to the availability of adequate photometric data (perhaps from Vera Rubin's LSST survey \cite{LSSTverarubin}) and an associated void catalog.  It also seems worth exploring whether adding spiral galaxies to the comparison sets would increase the statistical power of this method. 



\section*{Acknowledgements}
This work was supported by the US Department of Energy's Cosmic Frontiers program through award DE-SC0007881, and by Harvard University. We have used the Sloan Digital Sky Survey public  data archive. Funding for the Sloan Digital Sky Survey IV has been provided by the Alfred P. Sloan Foundation, the U.S. Department of Energy Office of Science, and the Participating Institutions. SDSS-IV acknowledges
support and resources from the Center for High-Performance Computing at
the University of Utah. The SDSS web site is www.sdss.org.

EP would like to thank his wife Shelbi Parker, for her invaluable support and proof-reading skills! EP and CS would also like to thank the Dillon Brout, Douglas Finkbeiner, and Elana Urbach for their discussions on the content of this paper. 

SDSS-IV is managed by the Astrophysical Research Consortium for the 
Participating Institutions of the SDSS Collaboration including the 
Brazilian Participation Group, the Carnegie Institution for Science, 
Carnegie Mellon University, the Chilean Participation Group, the French Participation Group, Harvard-Smithsonian Center for Astrophysics, 
Instituto de Astrof\'isica de Canarias, The Johns Hopkins University, Kavli Institute for the Physics and Mathematics of the Universe (IPMU) / 
University of Tokyo, the Korean Participation Group, Lawrence Berkeley National Laboratory, 
Leibniz Institut f\"ur Astrophysik Potsdam (AIP),  
Max-Planck-Institut f\"ur Astronomie (MPIA Heidelberg), 
Max-Planck-Institut f\"ur Astrophysik (MPA Garching), 
Max-Planck-Institut f\"ur Extraterrestrische Physik (MPE), 
National Astronomical Observatories of China, New Mexico State University, 
New York University, University of Notre Dame, 
Observat\'ario Nacional / MCTI, The Ohio State University, 
Pennsylvania State University, Shanghai Astronomical Observatory, 
United Kingdom Participation Group,
Universidad Nacional Aut\'onoma de M\'exico, University of Arizona, 
University of Colorado Boulder, University of Oxford, University of Portsmouth, 
University of Utah, University of Virginia, University of Washington, University of Wisconsin, 
Vanderbilt University, and Yale University.

\section*{Data Availability}
We have used the public data archive of the SDSS survey SpAll catalog: \url{https://data.sdss.org/sas/dr17/sdss/spectro/redux/v5_13_2/}, and the void catalog of \url{https://zenodo.org/record/5834787#.Y5uiXi-B2J9}].



\bibliographystyle{mnras}
\bibliography{VoidGalaxiesPedersen} 



\newpage
\appendix
\onecolumn
\section{Color distributions}
\label{App:colordist}
\begin{figure}
    \centering
    \includegraphics[width=\columnwidth]{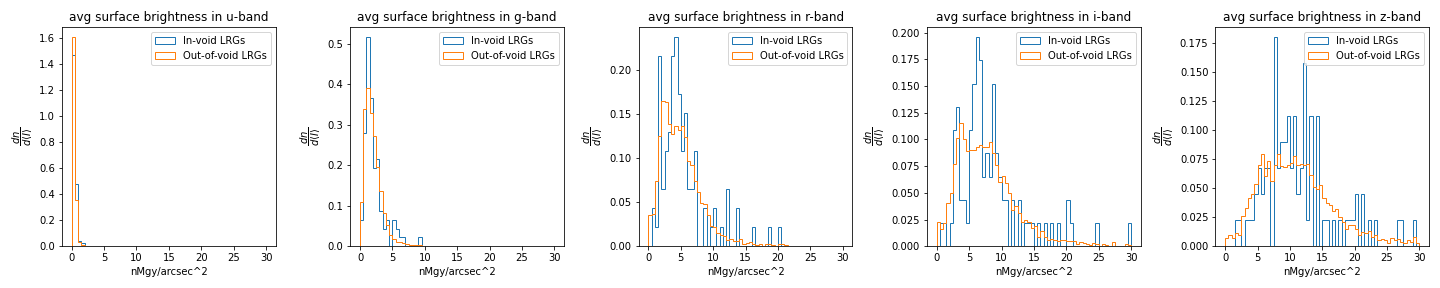}
    \caption{Surface brightness for each band. The bands are normalized individually, so the y-axis is not comparable between bands.}
    \label{fig:colors}
\end{figure}

\begin{figure}
    \centering
    \includegraphics[width=\columnwidth]{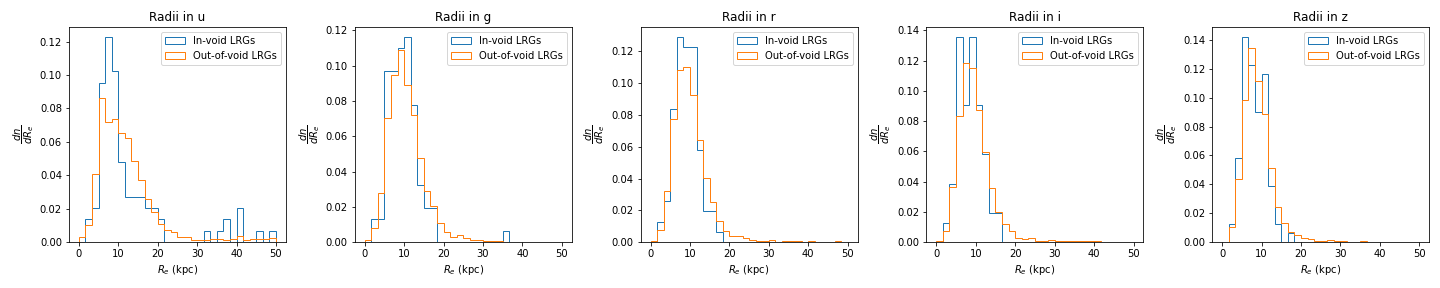}
    \caption{Radii estimated using the Petrosian angles $\theta_\text{petro}$} for each SDSS band. 
    \label{fig:radiis}
\end{figure}

Here we list the results using other colors used within SDSS. We illustrate the magnitudes in the other bands provided in SDSS, divided by the into the In-Void and Non-Void samples in Figure \ref{fig:colors}. While in Figure \ref{fig:radiis} we show the corresponding radii estimated as described in  Section \ref{sec:LRGprop}. For all but the u band we seem to get similar results for the $\alpha$ parameter introduced in Section \ref{sec:Galpha}, see Table (\ref{tab:alphacomp}). As can be seen all but the u band fall very closely together while the u band is still within the error obtained from the bootstrap. 

\begin{figure}
    \centering
    \includegraphics[width=\columnwidth]{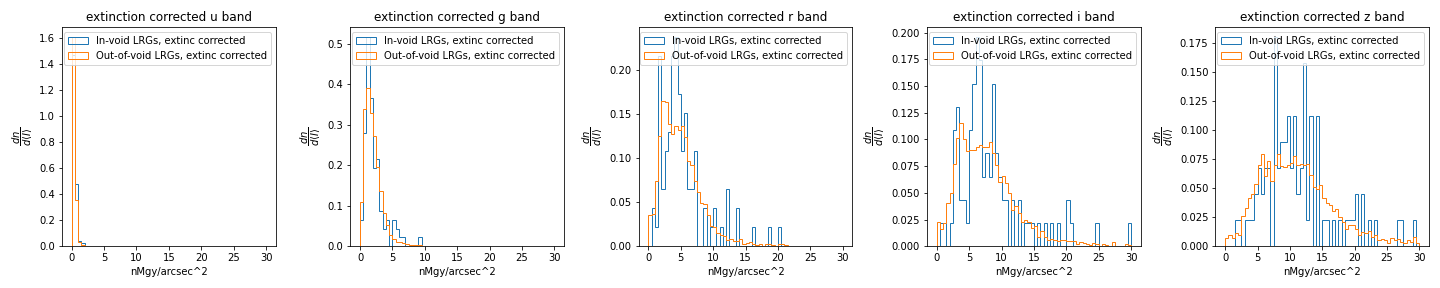}
    \caption{Extinction for each band, as histograms. The bands are normalized individually, so the y-axis is not comparable between bands.}
    \label{fig:extinction}
\end{figure}

\begin{table*}
    \centering
    \caption{Comparison of the $\alpha$ values obtained from each of the different bands. The uncertainty is obtained from the least square fit. The $r*$ is the value obtained using the restricted fundamental plane for the r magnitude.}
    \label{tab:alphacomp}
    \begin{tabular}{c|c}
\hline
band & $\alpha$ \\ 
\hline
u & -0.4871 $\pm$ 0.5134 \\ 
g & 0.0109 $\pm$ 0.4496 \\ 
r & -0.1072 $\pm$ 0.3010 \\ 
i & -0.1111 $\pm$ 0.2913 \\ 
z & -0.0235 $\pm$ 0.3476 \\ 
\hline
 $r*$ & 0.1312 $\pm$ 0.3844 \\ 
\hline
    \end{tabular}
    
\end{table*}

\bsp	
\label{lastpage}
\end{document}